\documentclass[aps,prb,reprint,amsmath,amssymb,showpacs,
groupedaddress]{revtex4-1}

\usepackage{graphicx}
\usepackage{dcolumn}
\usepackage{bm}

\begin{document}


\title{Commensurability oscillations in the rf conductivity of unidirectional lateral superlattices: measurement of anisotropic conductivity by coplanar waveguide}


\author{Akira Endo}
\email[]{akrendo@issp.u-tokyo.ac.jp}
\affiliation{The Institute for Solid State Physics, The University of Tokyo, 5-1-5 Kashiwanoha, Kashiwa, Chiba 277-8581, Japan}

\author{Toshiyuki Kajioka}
\affiliation{The Institute for Solid State Physics, The University of Tokyo, 5-1-5 Kashiwanoha, Kashiwa, Chiba 277-8581, Japan}

\author{Yasuhiro Iye}
\affiliation{The Institute for Solid State Physics, The University of Tokyo, 5-1-5 Kashiwanoha, Kashiwa, Chiba 277-8581, Japan}


\date{\today}

\begin{abstract}
We have measured the rf magnetoconductivity of unidirectional lateral superlattices (ULSLs) by detecting the attenuation of microwave through a coplanar waveguide placed on the surface. ULSL samples with the principal axis of the modulation perpendicular (S$_\perp$) and parallel (S$_\|$) to the microwave electric field are examined. For low microwave power, we observe expected anisotropic behavior of the commensurability oscillations (CO), with CO in samples S$_\perp$ and S$_\|$ dominated by the diffusion and the collisional contributions, respectively. Amplitude modulation of the Shubnikov-de Haas oscillations is observed to be more prominent in sample S$_\|$. The difference between the two samples is washed out with the increase of the microwave power, letting the diffusion contribution govern the CO in both samples. The failure of the intended directional selectivity in the conductivity measured with high microwave power is interpreted in terms of large-angle electron-phonon scattering.  
\end{abstract}

\pacs{73.43.Qt, 73.23.-b, 73.50.Mx}

\maketitle


\section{Introduction \label{intro}}
Radio-frequency (rf) conductivity has proven to be an incisive probe to investigate a two-dimensional electron gas (2DEG) subjected to a quantizing magnetic field, \cite{Engel93} comprising a unique tool to detect the pinning modes of various electron-solid-like ground states as resonant peaks seen in the frequency dependence. The states thus observed encompass Wigner crystals at small fillings of the lowest Landau level \cite{Ye02,Chen04,Chen06} or in the close vicinity of integer \cite{Chen03,Lewis04W} and fractional \cite{Zhu10} fillings, a Skyrme crystal, \cite{Zhu10S} and bubble \cite{Lewis02,Lewis04,Lewis05} and stripe \cite{Sambandamurthy08,Zhu09} phases formed at the partial fillings of high ($N \geq 2$) Landau levels.
In these studies, rf conductivity is evaluated by measuring the power transmission $\mathcal{T}$ through a coplanar waveguide (CPW) \cite{Wen69} placed on the surface of the 2DEG wafer (see Fig.\ \ref{samples}(a)). Microwave attenuates by coupling with the 2DEG underneath the slots of the CPW (Fig.\ \ref{samples}(b)). Since the absorbed power increases with increasing longitudinal conductivity $\sigma_{\alpha \alpha}$ ($\alpha = x$, $y$) of the 2DEG, lower $\mathcal{T}$ signifies higher $\sigma_{\alpha \alpha}$ (see Eqs.\ (\ref{sigmaP}), (\ref{PT}) below). The component $\sigma_{\alpha \alpha}$ thus measured is naively expected to represent the conduction along the direction of the microwave electric field $E_\text{rf}$, perpendicular to the propagation direction of the microwave (see Fig.\ \ref{samples}(a)). Although the direction does not matter when the conductivity is isotropic within the 2DEG plane ($x$-$y$ plane), it becomes necessary to identify the direction $\alpha$ correctly in the measurement of states having anisotropic conductivity, as is the case with the stripe phase at the half fillings of high Landau levels. \cite{Sambandamurthy08,Zhu09}

In the present study, we measure, employing the CPW method, rf magnetoconductivity of unidirectional lateral superlattices (ULSLs), the systems deliberately made anisotropic by introducing one-dimensional periodic potential modulation, $V(x)=V_0 \cos(2 \pi x / a)$. (We henceforth define the direction of the principal axis of the modulation as the $x$ direction, regardless of the propagation direction of the microwave.) By measuring samples with known anisotropy, we can identify the component of the conductivity detected in the measurement.

It is well known that ULSLs exhibit magnetoresistance oscillations --- the commensurability oscillations (CO) --- originating from the commensurability between the period $a$ of the modulation and the cyclotron radius $R_c = \hbar k_F/(e B)$ with $k_F = \sqrt{2 \pi n_e}$ the Fermi wavenumber and $n_e$ the areal electron density. \cite{Weiss89} The commensurability manifests itself via two different routes: diffusion and collisional contributions. \cite{Peeters92} Although both contributions can be traced back to the same origin, the oscillations with the magnetic field of the Landau bandwidth,
\begin{equation}
V_B \simeq V_0 \frac{1}{\pi} \sqrt{\frac{a}{R_c}} \cos \left( \frac{2 \pi R_c}{a}-\frac{\pi}{4} \right)
\label{VB}
\end{equation}
they differ markedly in their anisotropy, the phase, and the amplitude. The diffusion contribution $\Delta \sigma_{yy}^\text{dif}$ arises from the drift velocity in the $y$ direction $v_{\text{d},y}$ due to the dispersion of the Landau band. Accordingly, only $\sigma_{yy}$ component ($\alpha = y$) exists, and $\Delta \sigma_{yy}^\text{dif}$ vanishes, leading to minima in $\sigma_{yy}$, at the flat band conditions,
\begin{equation}
\frac{2 R_\text{c}}{a}=n-\frac{1}{4}\hspace{10mm}(n=1,2,3,...),
\label{flatband}
\end{equation}
where the bandwidth Eq.\ (\ref{VB}) equals zero. The collisional contribution $\Delta \sigma_{\alpha \alpha}^\text{col}$, on the other hand, results from the oscillations of the density of states and is therefore isotropic (seen equally in both $\sigma_{xx}$ and $\sigma_{yy}$ components), and takes maxima at the flat band conditions. Thus, $\Delta \sigma_{yy}^\text{dif}$ and $\Delta \sigma_{\alpha \alpha}^\text{col}$ oscillate with the same frequency (periodic in $1/B$) but with the opposite phase. The oscillation amplitude is usually much larger for the diffusion contribution, letting $\Delta \sigma_{yy}^\text{dif}$ dominate the oscillations in the experimental configuration in which $\sigma_{yy}$ component can be detected. By examining the phase of CO (whether minima or maxima are observed at the flat band conditions), therefore, one can see whether or not the $\sigma_{yy}$ component is involved in the measured conductivity.

We study two ULSL samples, S$_\perp$ and S$_\|$, differing in the orientation of the modulation. From the CO observed in these samples, we will show that for low microwave power the measured conductivity reflects the direction of $E_\text{rf}$ as expected. With the increase of the microwave power, however, the component of the conductivity diverted from the direction of $E_\text{rf}$ becomes mixed in and the oscillations are dominated by $\Delta \sigma_{yy}^\text{dif}$ for both samples. 

We note in passing that the great volume of CO reported thus far have been observed in the resistivity. The present study represents, to the knowledge of the present authors, the first observation of CO directly in the conductivity. 

The paper is organized as follows. In Sec.\ \ref{exp}, details of the experimental setup and the method of the measurement are described. Experimentally obtained rf magnetoconductivity traces exhibiting CO and the Shubunikov-de Haas oscillations (SdHO) are presented in Sec.\ \ref{CO}, which reveal that the difference between samples S$_\perp$ and S$_\|$ fades away with the increase of the microwave power. The origin of the loss of the intended orientational selectivity in the conductivity measurements is discussed in Sec.\ \ref{ephsc}, followed by concluding remarks in Sec.\ \ref{conc}.


%

\section{Experimental details \label{exp}}
\begin{figure}[tb]
\includegraphics[width=8.6cm]{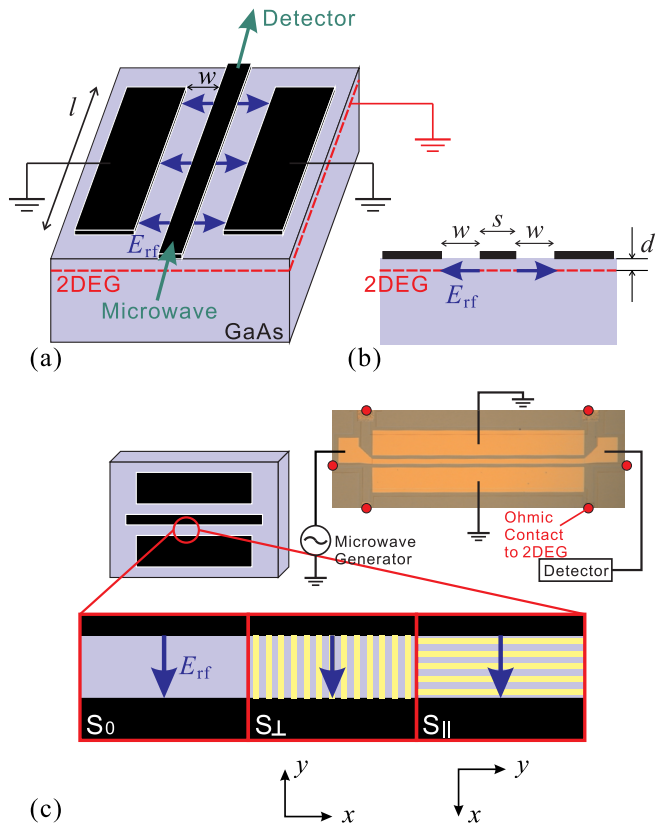}%
\caption{\label{samples} (Color online) (a) Schematic drawing of the device used in the measurements. (b) Cross section of (a). Microwave transmission through a coplanar waveguide (CPW, depicted by black plates) is measured. The microwave electric field $E_\text{rf}$ interacts with 2DEG beneath the two slots (width $w=28$ $\mu$m and length $l=1.6$ mm) between the CPW plates. The resulting microwave attenuation  reflects the conductivity of the 2DEG. (c) Illustration of the introduction of unidirectional modulation into the 2DEG in the slot region. In samples S$_\perp$ and S$_\|$, modulation is introduced with its principal axis ($x$ axis)  perpendicular and parallel to $E_\text{rf}$, respectively. Sample S$_0$ is for reference (without modulation). Inset: Optical micrograph of the whole device containing six Ohmic contacts for the resistivity measurements.}
\end{figure}

The devices for the measurements were fabricated from a conventional GaAs/AlGaAs 2DEG wafer  (mobility $\mu = 100$ m$^2$V$^{-1}$s$^{-1}$ and the electron density $n_e =3.8 \times 10 ^{15}$ m$^{-2}$) with the 2DEG plane residing at the depth $d = 60$ nm from the surface. As depicted in Fig.\ \ref{samples}, a metallic CPW having a center electrode (width $s = 40$ $\mu$m) flanked by two slots with the width $w = 28$ $\mu$m and the length $l = 1.6$ mm, designed to have the characteristic impedance $Z_0 = 50$ $\Omega$, was deposited on the surface using standard electron-beam (EB) lithography technique. Two ULSL samples, S$_\perp$ and S$_\|$ schematically illustrated in Fig.\ \ref{samples}(c), were prepared: in S$_\perp$ (S$_\|$), $x$ axis, namely the principal axis of the modulation, is set perpendicular (parallel) to the direction of the microwave electric field $E_\text{rf}$; the sample S$_\perp$ (S$_\|$) is originally designed to be sensitive to $\sigma_{yy}$ ($\sigma_{xx}$). We also prepared a device without modulation, S$_0$, and confirmed that oscillatory phenomena attributed to the modulation in Sec.\ \ref{CO} were absent in this sample. In ULSL samples, modulation was introduced by placing a grating of EB-resist on the slot regions of the surface. As in our previous studies, \cite{Endo00e,Endo05HH,Endo08ModSdH} the grating introduces potential modulation into the 2DEG via strain-induced piezoelectric effect. \cite{Skuras97}
\begin{figure}[tb]
\includegraphics[bbllx=0,bblly=140,bburx=715,bbury=540,width=8.6cm]{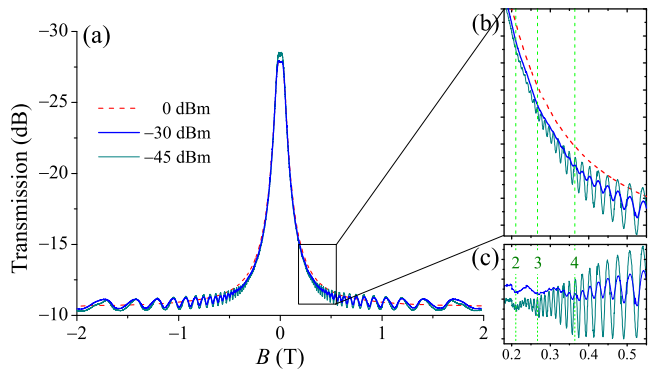}%
\caption{\label{rawdata} (Color online) (a) Microwave transmission $\mathcal{T}(B)$ for sample S$_\perp$ measured with a network analyzer, with the power (the output power of the network analyzer) $P_\text{NA}=$ $-$45, $-$30, and 0 dBm. The transmission is plotted in such a way that the upper side represents lower power transmission (higher attenuation, corresponding to higher conductivity $\sigma_{yy}$ of the 2DEG). $T_\text{bath}=$50 mK\@. (b) Enlarged plot of the portion surrounded by the rectangle in (a). (c) Oscillatory part of the traces for $-$30 and $-$45 dBm in (b), obtained by subtracting the featureless trace for 0 dBm. The flat band conditions, Eq.\ (\ref{flatband}), are indicated by vertical dashed lines in (b) and (c) along with the index $n$.}
\end{figure}

Measurements were performed in a top-loading dilution refrigerator using a probe equipped with rigid coaxial cables. Samples were immersed in the mixing chamber held at $T_\text{bath} = 50$ mK during the measurement. Microwave transmission $\mathcal{T}$ was measured with a network analyzer (Agilent Technology E5062) for various powers $P_\text{NA}$. We used the microwave frequency $\omega / (2 \pi) = 260$ or 300 MHz in the present study. No difference was observed between the two frequencies.
Typical magnetic-field dependence of the transmission $\mathcal{T}(B)$ is shown in Fig.\ \ref{rawdata}(a), taken for sample S$_\perp$ with three different values of $P_\text{NA}$ at the frequency 260 MHz. The power imparted to the electrons during the measurement inevitably heats up the 2DEG to the electron temperature $T_e$ higher than $T_\text{bath}$; the resultant $T_e$ is higher for higher microwave power. For traces taken with lower powers $P_\text{NA} = -30$ and $-45$ dBm and hence having lower $T_e$, SdHO are clearly observed. In the close-up shown in Fig.\ \ref{rawdata}(b), slower oscillations are also discernible, albeit very faintly. Oscillations are absent in the trace for the highest power $P_\text{NA} = 0$ dBm, apparently having considerably high temperature $T_e \agt 10$ K\@. More quantitative estimate of $T_e$ will be given below in Sec.\ \ref{ephsc}. The trace for $P_\text{NA} = 0$ dBm can be used as the slowly-varying background for the other traces. The oscillatory part thus extracted, shown in Fig.\ \ref{rawdata}(c), reveals that the slower oscillations have minima at the flat band conditions, Eq.\ (\ref{flatband}), and therefore is interpreted as the CO mainly due to the diffusion contribution $\Delta \sigma_{yy}^\text{dif}$. The dominance of the $\sigma_{yy}$ component is in accord with the expectation from the experimental setup.

The microwave with power $P_\text{NA}$ generated at the network analyzer (NA) attenuates during the propagation not only along the CPW but also along the entire circuitry (coaxial cables and connectors) that connects the sample and NA, before entering the NA again to be detected. For a quantitative account of the conductivity, therefore, it is necessary to extract the contribution due to the 2DEG from the measured transmission $\mathcal{T}(B)$. This is done by noting that a 2DEG does not absorb microwave when it is not conducting, $\sigma_{\alpha \alpha} = 0$. We single out the contribution attributable to the 2DEG $\Delta \mathcal{T}(B) = \mathcal{T}(B) - \mathcal{T}(\sigma_{\alpha \alpha} = 0)$ by subtracting $\mathcal{T}(\sigma_{\alpha \alpha} = 0)$ taken from the transmission at the quantum Hall plateau, e.g., $\mathcal{T}(B)$ at $B \sim 2$ T in Fig.\ \ref{rawdata}, assuming that the  transmission outside the CPW does not depend on $B$. By using the $\Delta \mathcal{T}(B)$ thus extracted and the standard distributed-circuit theory of the microwave transmission lines, \cite{Liao90,ChenPhD05} one obtains
\begin{equation}
\sigma_{\alpha \alpha} = - \frac{w}{2 l Z_0} \ln \mathcal{P} \sqrt {1 + \left( \frac{v_\text{ph}}{2 l \omega } \ln \mathcal{P} \right)^2 },
\label{sigmaP}
\end{equation}
where $\mathcal{P} = P_\text{out} / P_\text{in}$ is the ratio of the microwave power leaving $P_\text{out}$ and entering $P_\text{in}$ the CPW, with $\ln \mathcal{P} (< 0)$ related to $\Delta \mathcal{T}$ (in dB) as
\begin{equation}
\ln \mathcal{P} = \ln \left( \frac{P_\text{out}}{P_\text{in}} \right) = \left( \frac{\ln10}{10} \right) \Delta \mathcal{T},
\label{PT}
\end{equation}
and $v_\text{ph} = 1.12 \times 10^8$ m/s represents the phase velocity. For a large enough angular frequency $\omega$ and/or a small enough $|\ln \mathcal{P}|$ (corresponding to small enough $\sigma_{\alpha \alpha}$), Eq.\ (\ref{sigmaP}) can further be simplified as $\sigma_{\alpha \alpha} \simeq - w \ln \mathcal{P} / (2 l Z_0)$, which is usually a good approximation for measurements performed at high magnetic fields deep in the quantum Hall regime, where $\sigma_{\alpha \alpha}$ is small. \cite{Engel93,Ye02,Chen04,Chen06,Chen03,Lewis04W,Zhu10,Zhu10S,Lewis02,Lewis04,Lewis05,Sambandamurthy08,Zhu09,ChenPhD05} In the magnetic field range encompassed in the present study, however, $\sigma_{\alpha \alpha}$ remains relatively large. Furthermore, we had to resort to a relatively low frequency $\omega / (2 \pi) \leq 300$ MHz in order to retain a high s/n ratio that allows us to detect the small amplitude oscillations superposed on a large background. We therefore adhere to Eq.\ (\ref{sigmaP}) in evaluating the conductivity from $\Delta \mathcal{T}$.

\section{Commensurability oscillations and amplitude modulation of Shubnikov-de Haas oscillations\label{CO}}
\begin{figure}[tb]
\includegraphics[bbllx=25,bblly=85,bburx=515,bbury=740,width=8.6cm]{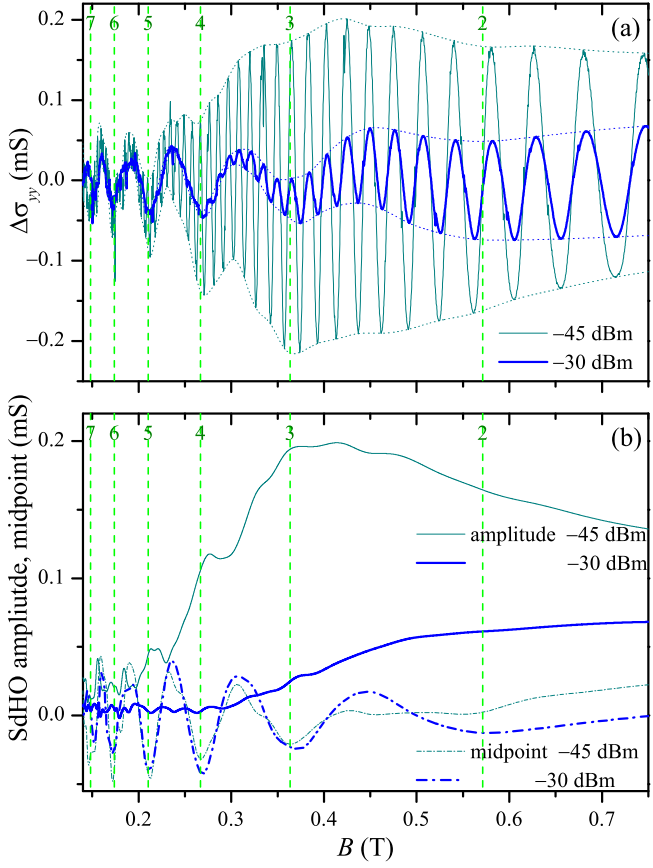}%
\caption{\label{osc-2B} (Color online) (a) Oscillatory part of the conductivity (at 260 MHz) for sample S$_\perp$ (having the configuration designed to measure the $\sigma_{yy}$ component) for $P_\text{NA}=$ $-$45 and $-$30 dBm. Upper and lower envelop curves, $\Delta \sigma_\text{UE}$ and $\Delta \sigma_\text{LE}$, are shown by dotted lines. (b) Amplitude (solid lines, $(\Delta \sigma_\text{UE}-\Delta \sigma_\text{LE})/2$) and midpoint (dot-dashed lines, $(\Delta \sigma_\text{UE}+\Delta \sigma_\text{LE})/2$) of the SdHO\@. The flat band conditions, Eq.\ (\ref{flatband}), are indicated by vertical dashed lines along with the index $n$.}
\end{figure}

In Fig.\ \ref{osc-2B}(a), we show the oscillatory part of the conductivity for sample S$_\perp$ for lower powers $P_\text{NA}=$ $-$45 and $-$30 dBm. The traces are obtained by first translating the transmission shown in Fig.\ \ref{rawdata}(a) into the conductivity using Eqs.\ (\ref{sigmaP}) and (\ref{PT}), and then by subtracting the slowly-varying background. \footnote{We used a carefully chosen function that monotonically decreases with $B$ without any inflection points. This turned out to work better than using the conductivity translated from the high power trace ($P_\text{NA} = 0$ dBm) in Fig.\ \ref{rawdata}(a) as a background. Note the remnant non-oscillatory component in Fig.\ \ref{rawdata}(c).} As can be seen from Eqs.\ (\ref{sigmaP}) and (\ref{PT}), the dependence of the conductivity on the transmission evolves from $\propto |\Delta \mathcal{T}|$ to $\propto |\Delta \mathcal{T}|^2$ with the increase of $|\Delta \mathcal{T}|$. In other words, the sensitivity of the transmission to the conductivity decreases with increasing $|\Delta \mathcal{T}|$, or equivalently with increasing conductivity. This accounts for the larger oscillation amplitude and the higher noise level seen in the lower magnetic field regime in Fig.\ \ref{osc-2B}(a) compared to Fig.\ \ref{rawdata}(c).

Two sets of the oscillations, SdHO and CO, are apparent in Fig.\ \ref{osc-2B}(a), with the former superposed on the latter. To gain more insight into those oscillations, the amplitude and the midpoint of the SdHO are drawn out as the half difference and the average, respectively, of the upper and lower envelop curves (plotted by dotted lines in Fig.\ \ref{osc-2B}(a)), and are shown in Fig.\ \ref{osc-2B}(b). The midpoint of SdHO, essentially corresponding to the CO, takes minima at the flat band conditions, consistent with the $\Delta \sigma_{yy}^\text{dif}$ as already mentioned in Sec.\ \ref{exp}. The amplitude of the SdHO also exhibits oscillations with the maxima occurring at the flat band conditions
(most obvious at $n=5$ and 4) for $P_\text{NA} = -45$ dBm. Amplitude modulation is less clear for $P_\text{NA} = -30$ dBm having much smaller SdH amplitude owing to higher $T_e$. 

The modulation of the SdHO amplitude in ULSLs derives from basically the same origin as CO, the diffusion and the collisional contributions. \cite{Peeters92,Endo08ModSdH} The diffusion contribution, namely the effect of $v_{\text{d},y}$ on the SdHO, affects only the $\sigma_{yy}$ component, and enhances the SdHO amplitude with the increment proportional to $V_B^2$, leading to the amplitude minima at the flat band conditions. The collisional contribution, by contrast, alters the SdHO amplitude isotropically. The SdHO amplitude is maximized at the flat band conditions, since there the broadening of the Landau levels (Landau bands) due to the dispersion vanishes. Again, the two contributions counteract each other. The maxima in the SdHO amplitude at the flat band conditions in Fig.\ \ref{osc-2B}(b) therefore signal the dominance of the collisional contribution. In a previous publication, \cite{Endo08ModSdH} the present authors have shown that the modulation of the SdHO amplitude in the magnetoresistance of ULSLs is dominated by the collisional contribution at low magnetic fields and that the diffusion contribution grows rapidly with the magnetic field and becomes the dominant contribution above $\sim$0.3 T. The behavior of the SdHO here is basically in line with that observed in the magnetoresistance. \cite{Endo08ModSdH} The peak of the SdHO amplitude is less apparent at $n=3$ compared to the peaks at $n=5$ and 4, which can be explained by the increase with the magnetic field of the relative importance of the diffusion contribution acting against the collisional contribution.
\begin{figure}[tb]
\includegraphics[bbllx=10,bblly=60,bburx=525,bbury=790,width=8.6cm]{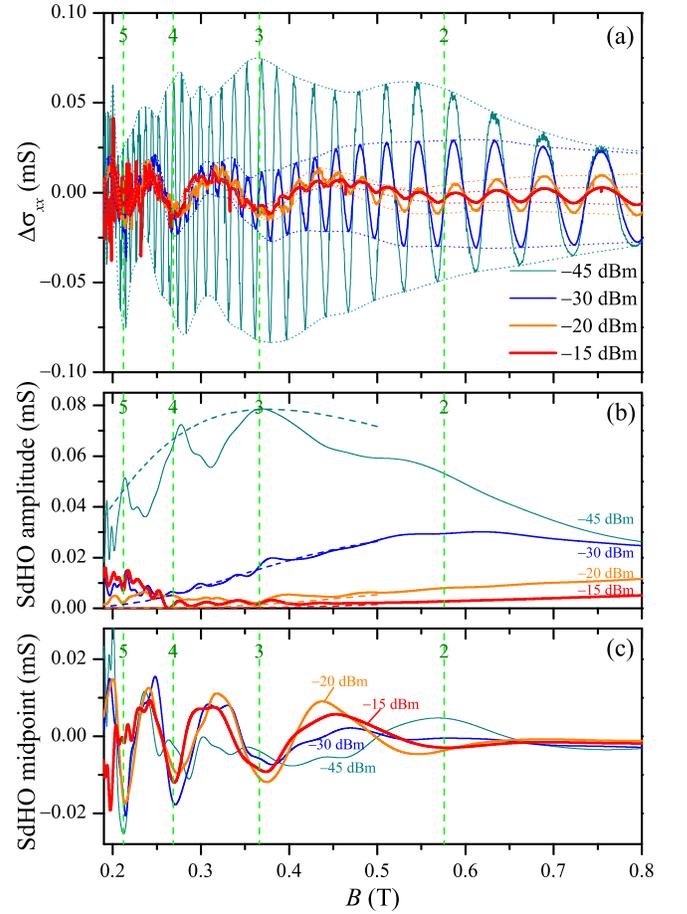}%
\caption{\label{osc-2C} (Color online) (a) Oscillatory part of the conductivity (at 300 MHz) for sample S$_\|$ (having the configuration designed to measure the $\sigma_{xx}$ component) for $P_\text{NA}=$ $-$45, $-$30, $-$20, and $-$15 dBm. Upper and lower envelop curves, $\Delta \sigma_\text{UE}$ and $\Delta \sigma_\text{LE}$, are shown by dotted lines. (b) Amplitude, $(\Delta \sigma_\text{UE}-\Delta \sigma_\text{LE})/2$, of the SdHO\@. Dashed curves show the fitting to Eq.\ (\ref{SdHamp}) of the values at the flat band conditions.  (c) Midpoint, $(\Delta \sigma_\text{UE}+\Delta \sigma_\text{LE})/2$, of the SdHO\@. The flat band conditions, Eq.\ (\ref{flatband}), are indicated by vertical dashed lines along with the index $n$.}
\end{figure}

We now turn to sample S$_\|$ having the modulation axis parallel to $E_\text{rf}$. First, we focus on the conductivity measured with the lowest power $P_\text{NA} = -45$ dBm (thinnest traces in Fig.\ \ref{osc-2C}). As shown in Fig.\ \ref{osc-2C}(a), the conductivity exhibits prominent SdHO. Amplitude modulation, with the maxima at the flat band conditions (Fig.\ \ref{osc-2C}(b)), is much more pronounced compared to that in sample S$_\perp$ (Fig.\ \ref{osc-2B}(b)). The SdHO midpoint plotted in Fig.\ \ref{osc-2C}(c) also exhibits maxima at the flat band conditions. These observations can be consistently understood by assuming that only the $\sigma_{xx}$ component is detected, as expected from the experimental configuration. The large amplitude modulation of the SdHO can be ascribed to the collisional contribution in the absence of the counteracting diffusion contribution. The behavior of the SdHO midpoint (representing CO) is also attributable to the collisional contribution, which would have been wiped out if the diffusion contribution were dominant.

With the increase of the microwave power $P_\text{NA}$, however, the diffusion contribution comes into play. As can be seen in Fig.\ \ref{osc-2C}(c), maxima at the flat band conditions observed in the SdHO midpoint turn into minima, implying that the collisional contribution in the CO is overridden by the diffusion contribution; the $\sigma_{yy}$ component dominated by the diffusion contribution $\Delta \sigma_{yy}^\text{dif}$ gets mingled in the conductivity measured with the setup aimed at probing $\sigma_{xx}$. \footnote{The amplitude modulation of the SdHO is also wiped out with the increase of $P_\text{NA}$, as seen in Fig.\ \ref{osc-2C}(b). Although this is mainly due to the dwindling of the SdHO itself with increasing $T_e$, it can also be affected by the intervention of the counteracting diffusion contribution.} We will consider the origin of the mixing of the component deflected from the direction of $E_\text{rf}$ in Sec.\ \ref{ephsc}.

\section{Orientational selectivity in the conductivity measurement \label{ephsc}}
\begin{figure}[tb]
\includegraphics[bbllx=65,bblly=25,bburx=740,bbury=550,width=8.6cm]{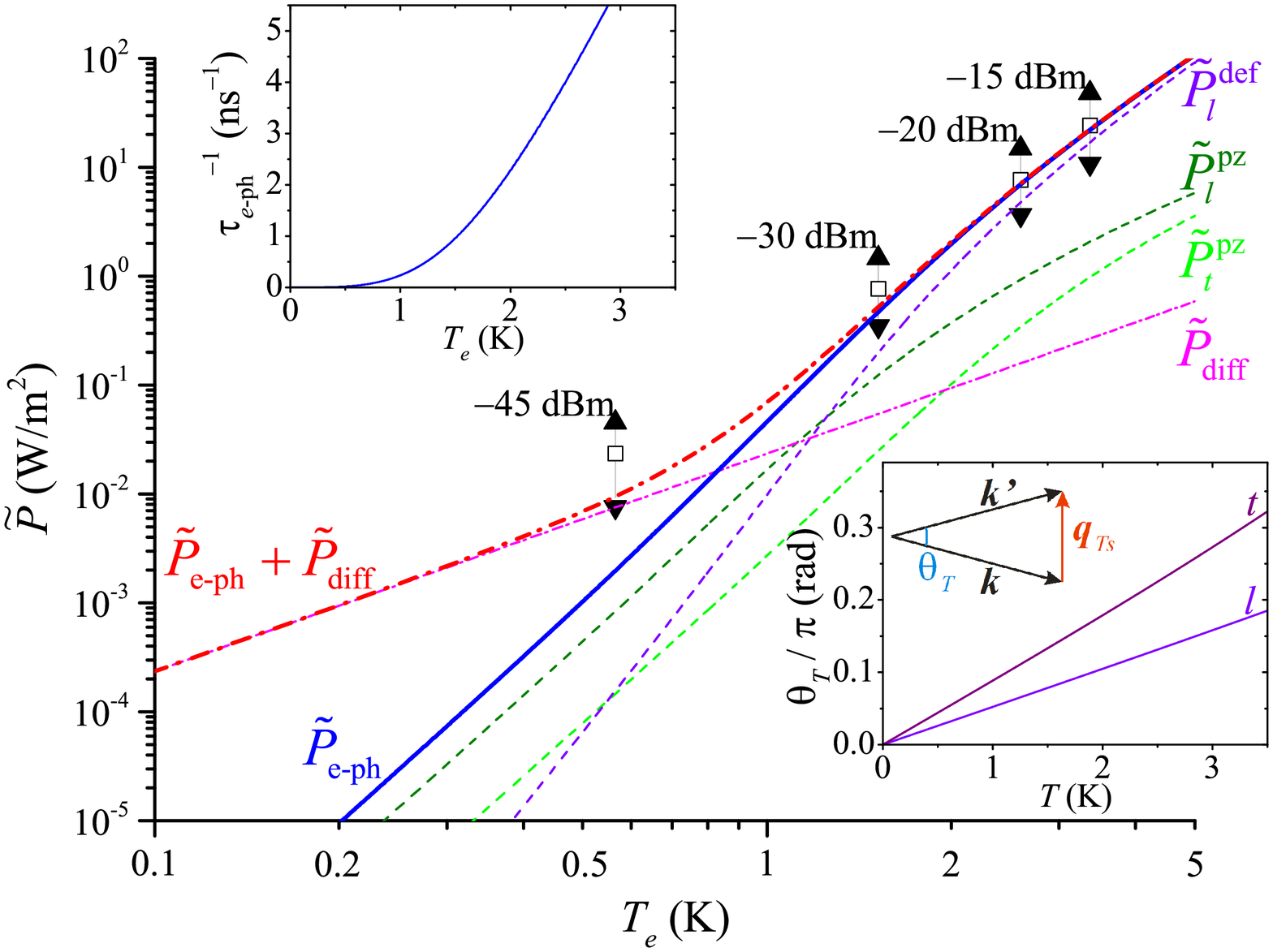}%
\caption{\label{TvsP} (Color online) Power absorbed by 2DEG per area $\widetilde{P}_\text{ab}$ vs\ electron temperature $T_e$ estimated from the SdHO amplitude, plotted in the log-log scale. The square, upward and downward triangles represent the average, the maximum, and the minimum power, respectively, absorbed in the magnetic field range 0.25 T $< B <$ 0.6 T\@. Lines represent calculated energy loss rates per area due to electron-phonon scattering vs $T_e$, assuming the lattice temperature $T_\text{L}$ to be equal to $T_\text{bath}=$ 50 mK: power loss by deformation potential coupling ($\widetilde{P}_l^\text{def}$), longitudinal and transverse piezoelectric potential coupling ($\widetilde{P}_l^\text{pz}$ and $\widetilde{P}_t^\text{pz}$) are plotted by thin dashed lines 
and their sum ($\widetilde{P}_\text{e-ph}$, Eq.\ (\ref{Ptot})) by a thick solid line. Roughly estimated energy loss rate by diffusion of electrons into the Ohmic contacts ($\widetilde{P}_\text{diff}$)%
, and the sum $\widetilde{P}_\text{e-ph}+\widetilde{P}_\text{diff}$ are also plotted by thin and thick dot-dashed lines, respectively. Upper inset: Electron-phonon scattering rate calculated by Eq.\ (\ref{scrate}). Lower inset: Scattering angle of the electron-phonon scattering by a typical phonon wavenumber $q_{Ts}$ $(s = l,t)$ at the temperature $T$ for longitudinal ($l$) and transverse ($t$) phonons. 
}
\end{figure}

In this section, we examine the role played by electron-phonon scattering in the process of electron heating by the microwave absorption. Since the electron-phonon scattering is a major source of large-angle scattering in 2DEGs, it is expected to significantly affect the directivity of the conductivity measurements delineated in Sec.\ \ref{CO}.

\subsection{Absorbed power versus electron temperature \label{expPabTe}}
As mentioned earlier, the microwave power $P_\text{ab} = P_\text{in} - P_\text{out}$ absorbed by the 2DEG lying underneath the slots of the CPW raises the local electron temperature $T_e$ above the lattice temperature $T_\text{L}$ (two-bath model). The heated 2DEG, in turn, transfers the power $P_\text{e-ph}$ to the phonons via electron-phonon scattering or $P_\text{diff}$ to the Ohmic contacts through diffusion. We assume in the present paper that both the lattice temperature $T_\text{L}$ and the temperature of the Ohmic contacts are the same as the temperature $T_\text{bath} = 50$ mK of the $^3$He-$^4$He mixture in which the sample is immersed. The power thus flowing out of the 2DEG increases with $T_e$, and $T_e$ settles to the value at which
\begin{equation} 
P_\text{e-ph}(T_e, T_\text{L}) + P_\text{diff}(T_e, T_\text{L}) = P_\text{ab}
\label{Pbalance}
\end{equation}
holds for a given $P_\text{ab}$. The relation Eq.\ (\ref{Pbalance}) has been used to scrutinize the details of the electron-phonon interaction embodied in the term $P_\text{e-ph}(T_e, T_\text{L})$, \cite{Price82,Karpus86,*Karpus87,*Karpus88} by examining the relation between the known amount of $P_\text{ab}$ (introduced to the sample by the Joule heating) and the $T_e$ measured with various methods of thermometry, including the self resistance, \cite{Wennberg86,Mittal94,Mittal96,Mittal96N,Chow97} the amplitude of the SdHO, \cite{Hirakawa86,Ma91,Chow96} thermopower of the ballistic constriction, \cite{Appleyard98,Proetto91} and the slope of the Hall resistance between two adjacent quantum Hall states. \cite{Chow96} 

Following these studies, we plot in Fig.\ \ref{TvsP} the power per area $\widetilde{P}_\text{ab} = P_\text{ab}/(2 w l)$ absorbed from the microwave versus $T_e$ for sample S$_\|$ taken from the set of the data shown in Fig.\ \ref{osc-2C}. Here and in what follows, we use the tilde $\widetilde{P}$ and the hat $\hat{P}$ to designate the power per area and the power per electron, respectively, related by $\widetilde{P} = \hat{P} \cdot n_e$. The electron temperature $T_e$ is deduced by fitting the amplitude of SdHO at the flat band conditions (where the effect of $V(x)$ on the amplitude vanishes) to the standard formula (an approximation valid for $\mu B \gg 1$), \cite{AndoRev82,Isihara86,Endo09FR}
\begin{equation}
\Delta \sigma_\text{SdHOamp} \simeq \frac{\sigma_0}{1+(\mu B)^2} \exp \left( -\frac{\pi}{\mu_Q B} \right) A \left( 2 \pi^2 \frac{k_\text{B} T_e}{\hbar \omega_c} \right),
\label{SdHamp}
\end{equation}
with $A(X)=X / \sinh(X)$, $\sigma_0 = n_e e \mu$ the conductivity at $B=0$, $\mu_Q$ the quantum mobility, \footnote{We used $\mu_Q = 6.15$ m$^2$V$^{-1}$s$^{-1}$ deduced from the SdHO in the magnetoresistivity measured separately with a Hall bar device fabricated from the same 2DEG wafer as that used for samples S$_\perp$ and S$_\protect{\|}$.} and $\omega_c = eB/m^*$ the cyclotron angular frequency with $m^*$ the effective mass. The results of the fitting are shown in Fig.\ \ref{osc-2C}(b) by dashed lines. The power absorbed by the 2DEG, $P_\text{ab} = P_\text{in} (1-\mathcal{P})$, is calculated from $\mathcal{P}$ obtained by the measured $\mathcal{T}(B)$ and Eq.\ (\ref{PT}), and $P_\text{in}$ translated from $P_\text{NA}$ by $P_\text{in} = P_\text{NA} \cdot 10^{(\mathcal{T}_0/20)}$ with $\mathcal{T}_0 \equiv \mathcal{T}(\sigma_{\alpha \alpha} = 0)$ the transmission (in dB) through the entire circuit in the absence of the attenuation due to the 2DEG\@. \footnote{Here, we used the fact that the incoming and outgoing coaxial cable circuitry is symmetric.} The transmission $\mathcal{T}(B)$ hence $P_\text{ab}$ varies with $B$ resulting from the variation in $\sigma_{\alpha \alpha}$, and the maximum and minimum values of $\widetilde{P}_\text{ab}$ in the range 0.25 T $< B <$ 0.6 T are indicated in the plots.

In Fig.\ \ref{TvsP}, the plots are compared with calculated curves of $\widetilde{P}_\text{e-ph}(T_e, T_\text{L})$ and $\widetilde{P}_\text{diff}(T_e, T_\text{L})$. For the electron-phonon scattering, we basically follow a theory by Price \cite{Price82} (with the correction of a factor of 2 error pointed out in Ref.\ \onlinecite{Mittal96N}). The formulas used in the calculation are outlined in the subsequent subsection Sec.\ \ref{ephtheory} with the details and subtleties of the approximations employed.

\subsection{Calculation of power loss by electron-phonon scattering as a function of electron temperature \label{ephtheory}}
The total electron-phonon energy transfer rate (per area),
\begin{equation}
\widetilde{P}_\text{e-ph}=\widetilde{P}_l^\text{def}+\widetilde{P}_l^\text{pz}+2 \widetilde{P}_t^\text{pz},
\label{Ptot}
\end{equation}
is composed of deformation-potential coupling and piezoelectric coupling contributions (denoted by the superscript $r =$ def and pz, respectively) with the longitudinal mode (the subscript $s = l$) and two branches of the transverse modes ($s = t$) for the latter coupling. In the strong screening condition (to be discussed below), the constituent components in the $P_\text{e-ph}$ can be written down (per electron) as \cite{Price82}
\begin{equation}
\hat{P}_s^r (T_e,T_\text{L}) = \hat{\Pi} _s^r (T_e) - \hat{\Pi} _s^r (T_\text{L}),
\label{PowerLosspere}
\end{equation}
with
\begin{equation}
\hat{\Pi}_l^\text{def}(T) =
\frac{1}{\varepsilon_\text{F} k_\text{F}} \frac{D^2 m^* v_l {a_\text{B}^*}^2}{16\pi \rho}\left( \frac{k_\text{B} T}{\hbar v_l} \right)^7 G_l ^{\text{def}} \left( \kappa _\text{F} \right),
\label{Pidefl}
\end{equation}
and
\begin{equation}
\hat{\Pi}_s^\text{pz}(T) =
\frac{1}{\varepsilon_\text{F} k_\text{F}}\frac{(eh_{14})^2 m^* v_s {a_\text{B} ^*}^2}{16\pi \rho}\left( \frac{k_\text{B} T}{\hbar v_s} \right)^5 G_s ^{\text{pz}} \left( \kappa_\text{F} \right),
\label{Pipzs}
\end{equation}
where $\varepsilon_\text{F}$ and $k_\text{F}$ represent the Fermi energy and wavenumber, respectively. We used in the calculation the following well-known material parameters for GaAs \cite{Adachi85,Lyo88,*Lyo89,Endo05HH}: the deformation potential $D = -9.3$ eV, the piezoelectric constant $h_{14} = 1.2\times10^9$ V m$^{-1}$, the mass density $\rho = 5.3$ g cm$^{-3}$, the effective Bohr radius $a_\text{B}^* =10.4$ nm, the effective mass $m^* =0.067m_e$ (with $m_e$ the bare electron mass), and the longitudinal and transverse sound velocities $v_l = 5.14\times10^3$ m s$^{-1}$ and $v_t = 3.04\times10^3$ m s$^{-1}$, respectively. The dimensionless function $G_s ^r (\kappa_\text{F})$ is written as \cite{Price82}
\begin{eqnarray}
G_s ^r (\kappa_\text{F}) \equiv &
\displaystyle{\frac{1}{\pi} \int_{-\infty}^\infty  d\zeta \left| F (q_{Ts} \zeta ) \right|^2 \times \hspace{25mm}} \nonumber \\
 & \displaystyle{\int_0^{\kappa_\text{F}} \! \! \! \! \! \frac{d\xi}{\sqrt{1 - (\xi / \kappa_\text{F})^2}} \frac{g_s^r (\xi , \zeta)}{e^{\sqrt{\xi ^2  + \zeta ^2 } }  - 1} \frac{1}{H^2( q_{Ts} \xi )} },
\label{Gsr}
\end{eqnarray}
using the form factor, 
\begin{equation}
F( q_z ) = \int {dz |\Phi (z)| ^2} e^{iq_z z},
\label{FormFactor}
\end{equation}
and a function related to the screening,
\begin{equation}
H( q_{\|} ) = \iint {dz_1 dz_2 |\Phi (z_1)| ^2  |\Phi (z_2)| ^2} e^{-q_{\|} |z_1 - z_2|},
\label{ScrnFactor}
\end{equation}
which are the functions of the components of the phonon wavevector perpendicular $q_z$ and parallel $q_\|$ to the 2DEG plane, respectively, and $\Phi (z)$ represents the envelope of the 2DEG wavefunction in the $z$ direction. 
In Eq.\ (\ref{Gsr}), we used $q_{Ts} \equiv k_\text{B} T / \hbar v_s$ to normalize 2$k_\text{F}$ and the components of the phonon wavevector: $\kappa_\text{F} \equiv 2 k_\text{F}/q_{Ts}$, $\xi \equiv q_\|/q_{Ts}$, and $\zeta \equiv q_z/q_{Ts}$.
The kernel $g_s^r (\xi , \zeta)$ in the integral Eq.\ (\ref{Gsr}) is given by \cite{Price82}
\begin{equation}
g_l ^\text{def} (\xi , \zeta) \equiv \xi ^2 (\xi ^2 + \zeta ^2) ^{3/2},
\label{gdefl}
\end{equation}
\begin{equation}
g_l ^\text{pz} (\xi , \zeta) \equiv 
\frac{ 9 \xi ^6 \zeta ^2 }{2 (\xi ^2 + \zeta ^2) ^{5/2} },
\label{gpzl}
\end{equation}
and
\begin{equation}
g_t ^\text{pz} (\xi , \zeta) \equiv 
\frac{ 8 \xi ^4 \zeta ^4 + \xi ^8}{4 (\xi ^2 + \zeta ^2) ^{5/2} }.
\label{gpzt}
\end{equation}
In an ideal 2DEG $\Phi (z) = \delta (z)$, we have $|F(q_z)| = H(q_{\|}) =1$. Since $q_{Ts}$ is smaller than the inverse of the rms thickness $\sim$5 nm of our 2DEG \cite{Endo05MA} in the temperature range $T_\text{L} < T_e \alt 5$ K encompassed in the present study, we can replace $|F(q_z)|$ and $H(q_{\|})$ in Eq.\ (\ref{Gsr}) by unity to a good approximation. Note that smaller values of $\xi$ and $\zeta$ weigh more in Eq.\ (\ref{Gsr}) owing to the exponential factor in the denominator $[\exp (\sqrt{\xi ^2 + \zeta ^2})-1]^{-1}$, which derives from the Bose-Einstein distribution of the phonons, $N_s(q,T) = [\exp(q/q_{Ts})-1]^{-1}$. The smallness of $q_{Ts}$ also leads to a small value of $a_\text{B}^* q_{\|}$, allowing the static dielectric function $\epsilon (q_{\|}) = 1 + 2 H(q_{\|})/(a_\text{B}^* q_{\|})$ to be approximated by $\epsilon (q_{\|}) \simeq 2 H(q_{\|})/(a_\text{B}^* q_{\|})$. This is the strong screening condition used in the derivation of Eqs.\ (\ref{Pidefl}) and (\ref{Pipzs}). Using these approximations, we numerically evaluate Eq.\ (\ref{PowerLosspere}) and translate the values of $\hat{P}_s^r(T_e,T_\text{L})$ ($r=\text{def, pz}$, $s = l$, $t$) to the power losses per area by 
\begin{equation}
\widetilde{P}_s^r(T_e,T_\text{L}) = \hat{P}_s^r(T_e,T_\text{L}) \cdot n_e
\label{PowerLossperA}
\end{equation}
and plot them along with their sum Eq.\ (\ref{Ptot}) in Fig.\ \ref{TvsP} against $T_e$. From the energy loss rate, we can also calculate the electron-phonon scattering rate by \cite{Mittal96N}
\begin{equation}
\frac{1}{\tau _\text{e-ph}} = \frac{1}{C_e} \frac{\partial \widetilde{P}_\text{e-ph}}{\partial T_e},
\label{scrate}
\end{equation}
where $C_e  = (\pi ^2 k_{\text{B}} ^2 / 3) D ( \varepsilon _\text{F})T_e = e^2 L_0 D ( \varepsilon _\text{F})T_e $ represents the electron specific heat (per area), with $D(\varepsilon)$ the density of states and $L_0  = \pi ^2 k_\text{B} ^2 / (3e^2) = 2.44 \times 10^{ - 8}$ W$\Omega$K$^{-2}$ the Lorenz number. The scattering rate $\tau_\text{e-ph}^{-1}$ thus calculated is plotted in the left inset to Fig.\ \ref{TvsP}.

\subsection{Dominant power-loss mechanism and orientational selectivity in the conductivity measurement \label{cmpexpth}}
We can see in Fig.\ \ref{TvsP} that, in the measurements carried out with higher microwave powers $P_\text{NA} = -30$, $-20$, and $-15$ dBm, the predominant power-loss mechanism responsible for determining $T_e$ is the electron-phonon scattering. By contrast, the electron-phonon scattering plays only minor role for the lowest power $P_\text{NA} = -45$ dBm. In this case, the energy is transferred mainly through the diffusion of the electrons into the Ohmic contact.
We roughly estimate the energy loss rate due to the diffusion using the formula \cite{Mittal96N}
\begin{equation}
P_\text{diff}  = \frac{W}{L} \frac{L_0 \sigma_{xx} }{2} (T_e ^2  - T_\text{L} ^2),
\label{diff}
\end{equation}
with $W$ and $L$ representing the (effective) width and length, respectively, of the path connecting the heated area to the Ohmic contact pads. 
In Eq.\ (\ref{diff}), the Wiedemann-Franz law $\kappa_{xx} = L_0 \sigma_{xx} T$ is employed to deduce the thermal conductivity $\kappa_{xx}$ from the electric conductivity $\sigma_{xx}$. The calculated $\widetilde{P}_\text{diff} = P_\text{diff}/(2wl)$ shown in Fig.\ \ref{TvsP} at least qualitatively explains the deviation at $P_\text{NA} = -45$ dBm from the trend  $\widetilde{P}_\text{ab}(T_e) \simeq \widetilde{P}_\text{e-ph}(T_e)$ followed by the higher power measurements; \footnote{To apply the formula Eq.\ (\ref{diff}) intended for a rectangular sample to our complicatedly shaped sample (see the inset of Fig.\ \ref{samples}(c)), we resorted to drastic approximation: we used the values of $W/L$ taken from the thinnest part of the legs leading to the six Ohmic contact pads, and the value of $\sigma_{xx}$ at $B=0$, hoping that the cancellation of the under- and overestimating nature of the two treatments yields the right order of magnitude. The ambiguity in the quantitative estimate of $P_\text{diff}$ does not affect the main conclusion of the present paper.} it is obvious that $P_\text{diff} \propto T_e^2$, having the lowest power 2 for the temperature dependence, becomes the dominant contribution at low enough temperatures. 

Figure \ref{TvsP} combined with Fig.\ \ref{osc-2C}(c) reveals that when the microwave power injected into the 2DEG from the CPW is dissipated mainly via electron-phonon scattering, the measured conductivity loses the information of the direction of $E_\text{rf}$.
The intended orientationally-selective measurement is possible only for a weaker microwave power, where the electron-phonon scattering accounts for only a minor portion of the power dissipation. 
It is natural to relate the loss of the directional selectivity to the large scattering angle the electrons undergo in the electron-phonon scattering events. Scattering angle due to the typical phonon wavenumber $q_{Ts}$ at the temperature $T$ (with $s = l$, $t$), $\theta_T = 2 \arcsin(q_{Ts}/2k_\text{F})$, is plotted in the right inset of Fig.\ \ref{TvsP}. It shows that, for high $P_\text{NA}$ with high $T_e$, the scattering angle due to the electron-phonon scattering \footnote{The scattering angle is primarily determined by the phonons at the temperature $T_e$, not the lattice temperature $T_\text{L}$. Note the dominance of the first term in Eq.\ (\ref{PowerLosspere}), which contains the Bose-Einstein distribution with the temperature $T_e$, $N_s(q,T_e)$ (see Eqs.\ (\ref{PowerLosspere})--(\ref{Gsr})).} can become much larger than the average scattering angle due to the impurity scattering, $\theta_\text{imp} \sim \arccos(1-\mu_Q/\mu) \sim 0.1\pi$ rad. Note that $\theta_\text{imp}$ is small in modulation doped GaAs/AlGaAs 2DEGs, in which remote ionized donors constitute the major source of the impurity scattering. \cite{Davies98B}  

It is interesting to point out that, even when the electron-phonon scattering is the main power-loss mechanism, the scattering rate $\tau_\text{e-ph}^{-1}$ is still orders of magnitude smaller than the cyclotron frequency (100 ns$^{-1}$ $ < \omega_c / (2 \pi) <$ 250 ns$^{-1}$ in 0.25 T $< B <$ 0.6 T) or the impurity scattering rate, $\tau_Q^{-1} = e/(m^* \mu_Q) \simeq 430$ ns$^{-1}$; electrons experience numbers of cyclotron revolutions and impurity scatterings, without being disturbed by large-angle scatterings, before passing on the energy gained from the CPW to the phonons. Nevertheless, the direction of $E_\text{rf}$ is not reflected in the measured conductivity when the electron-phonon scattering is the major route of the power loss. This implies that the whole process, from the initial acceleration of the electrons by $E_\text{rf}$ to the dumping of the energy thus obtained, should be devoid of large angle scatterings in order for the attenuation of the microwave through the 2DEG to be sensitive only to the component of the conductivity parallel to $E_\text{rf}$. 

\section{Conclusions \label{conc}}
By measuring the attenuation of microwave through a coplanar waveguide (CPW), we have observed commensurability oscillations (CO) in the rf magnetoconductivity of unidirectional lateral superlattices (ULSLs) embedded in the slots of the CPW\@. Using the phase of the CO as an indicator, we can examine whether or not the component of the conductivity parallel to the rf electric field $E_\text{rf}$ can be probed selectively. For low microwave power, we find that the component parallel to $E_\text{rf}$ is detected as expected. With the increase of the microwave power, however, the CO becomes dominated, regardless of direction the principal axis ($x$ axis) of the modulation, by the diffusion contribution which is contained only in $\sigma_{yy}$. In the latter power range, the microwave power absorbed by the 2DEG raises the electron temperature $T_e$ so high ($T_e \agt 1.5$ K) above the lattice temperature $T_\text{L} = 50$ mK that the power is dissipated mainly through the electron-phonon scattering. We attribute the mixing of the component deflected from the direction of $E_\text{rf}$ to the large scattering angle brought in by the electron-phonon scattering. In order to preserve the intended orientation in the conductivity measurement, the microwave power should be kept low enough to prevent the electron-phonon scattering from becoming the major power loss channel.


%



\begin{acknowledgments}
This work was supported in part by Grant-in-Aid for Scientific Research (A) (18204029), (B) (20340101), and (C) (22560004) from the Ministry of Education, Culture, Sports, Science and Technology (MEXT).
\end{acknowledgments}

\bibliography{ourpps,lsls,ninehlvs,wc,thermo,twodeg}

\end{document}